\documentclass[%
 aip,
rsi,%
 amsmath,amssymb,
 reprint,%
]{revtex4-1}

\usepackage{graphicx}
\usepackage{dcolumn}
\usepackage{bm}

\begin{document}

\preprint{AIP}

\title[]{A surface-hopping method for semiclassical calculations of cross sections for radiative
   association with electronic transitions}

\author{P\'eter Szab\'o}
 \email{peter.szabo@ltu.se}
 \affiliation{Applied Physics, Division of Materials Science, Department of Engineering Science and Mathematics,
              Lule\aa \space University of Technology, 97187 Lule\aa, Sweden}

\author{Magnus Gustafsson}%
 \affiliation{Applied Physics, Division of Materials Science, Department of Engineering Science and Mathematics,
              Lule\aa \space University of Technology, 97187 Lule\aa, Sweden}


\date{\today}

\begin{abstract}
 A semicalssical method based on surface-hopping techniques is developed
 to model the dynamics of radiative association with electronic transitions
 in arbitrary polyatomic systems. It can be proven that our method
 is an extension of the established semiclassical formula used
 in the characterization of diatomic molecule-formation.
 Our model is tested for diatomic molecules. It gives the same cross sections as the former
 semiclassical formula, but contrary to the former method it allows us to follow
 the fate of the trajectories after the emission of a photon. This means that
 we can characterize the rovibrational states of the stabilized molecules:
 using semiclassial quantization we can obtain quantum state resolved
 cross sections or emission spectra for the radiative association process.
 The calculated semiclassical state resolved spectra show good agreement
 with the result of quantum mechanical perturbation theory.
 Furthermore our surface-hopping model is not only applicable for
 the description of radiative association but it can be use for
 semiclassical characterization of any molecular process
 where spontaneous emission occurs.

\end{abstract}

\maketitle


\section{Introduction}

 Radiative association is one of the many important processes \cite{Babb_Kirby_98} that contribute
 to the molecule production in dust-poor regions of interstellar space
 where there are few competing three-body collisions. Radiative association is hard
 to study under laboratory conditions due to the small cross sections of the process \cite{Gerlich_92_cr_92_1509}.
 Experimental studies on radiative rate coefficients have been carried out so far
 only for a few ionic systems using ion traps or ion cyclotron resonance apparatus \cite{Gerlich_92_cr_92_1509}.
 So far there is no experimental result for the formation of neutral molecules through
 radiative association. Theoretical modeling of radiative association is highly
 desirable due to the lack of experimental data. In the case of neutral species it
 is the only possible way to obtain the rate coefficients \cite{Rev_theo}. Good quality potential energy
 and dipole surfaces are nowadays available from accurate
 ab initio calculations \cite{Czako_1,Czako_2,Jensen,Tennyison}. Adequate dynamical methods based on ab initio
 data can provide reliable rate coefficients that can be utilized for kinetic modeling
 of molecule production in the interstellar medium \cite{OHO}.

 Most of the previous dynamical calculations have focused on diatomic systems 
 \cite{Rev_theo,Antipov_09_jcp,Antipov_11_jcp,Barinovs_05_ApJ_636_923,Bennett_03_MonNotRAstronSoc_341_361,Dickinson_05_JPhysB_38_4329,CO_paper1,HF_paper,CO_paper2,Singh_00_ApJ_537_261}.
 There are only a few dynamical studies where the
 radiative association of triatomic molecules have been considered \cite{Mrugala_2003,Mrugala_2005,Mrugala_2015,Stoecklin,Ayouz,Talbi}.
 Radiative association is not only difficult to study experimentally,
 but also theoretically, despite the available global potential energy
 and dipole surfaces\cite{Rev_theo}. There are several reasons which can explain the small
 number of dynamical studies on polyatomic systems: i) the quantum mechanical treatment
 of radiative association is a major challenge because we need both bound and unbound states \cite{Mrugala_2003,Mrugala_2005,Mrugala_2015,Stoecklin,Ayouz};
 ii.) so far there has been no available semiclassical
 method which can describe the radiative association of polyatomic molecules
 with electronic transitions in full dimension; iii) a classical model for
 the calculation of radiative association in absence of electronic transitions
 has been recently developed\cite{Classical}, but not yet implemented for polyatomic systems. The semiclassical methods are based
 on classical trajectories and therefore do not contain the resonance structure
 and the tunneling in the cross sections. In spite of the lack of resonance
 and tunneling effects, trajectory based methods can provide reasonable
 results in many cases \cite{Szabo_1,Szabo_2,HBrOH}. The extension of the theoretical studies for
 polyatomic systems is highly desirable. The bigger the reactants are the more
 probable radiative association is \cite{Herbst_1,Herbst_2,Herbst_3}. Semiclassical and classical methods are the only
 feasible way to the theoretical modeling of the dynamics of systems which contain
 more than 3-4 atoms.

 The purpose of the current work is to present a new semiclassical dynamical method for the
 modeling of radiative association with electronic transitions. We use the surface
 hopping method extended with Fermi's golden rule which ensures a simple way
 to the treatment of dynamical coupling of molecular states with electromagnetic fields.
 The surface-hopping methodology allows the study of radiative association
 with electronic transitions to be extended to arbitrary polyatomic system.

 The surface-hopping method is widely used in the modeling of nonadiabatic molecular processes 
 \cite{Thiel,Barbatti,SHARC,JADE,Subotnik_rev,Subotnik_1,Subotnik_2,FISH,Mitric_1,Mitric_2,Mitric_3,Mitric_4,Gonzalez_1,Gonzalez_2,Thiel_2,Thiel_3,Barbatti_2}.
 The idea of this method originally put forward by
 Bjerre and Nikitin \cite{Surfhopp_original_1} and later advanced by others 
 \cite{Surfhopp_original_2,Surfhopp_original_3,Surfhopp_original_4,Surfhopp_original_5,Surfhopp_original_6}.
 The motion of nuclei are described by classical mechanics but the time evolution of
 the molecular electronic states is treated quantum mechanically. Classical trajectories
 can take place only on one potential energy surface at a time. Therefore we need
 a stochastic algorithm that allows the change between potential energy surfaces
 during the propagation.  The branching of the populations
 due to the dynamical coupling is simulated with this stochastic change in electronic states.
 Generally the wavefunction of the adiabatic electronic states is used as basis set
 for the representation of the time-dependent wavefunction. The population of
 the electronic states (the hopping probability) can be calculated in each time step
 from the solution of the time-dependent Schr\"odinger equation. The ensemble  of the
 independent trajectories extended with stochastic hopping can provide a reliable
 semiclassical approximation for the simulation of nonadiabatic dynamics.

 The surface-hopping method was extended in the last decade for the treatment of arbitrary coupling \cite{FISH,SHARC,Thiel}.
 Field-induced couplings were the subject of more studies where the absorption spectra \cite{Subotnik_rev,Subotnik_1,Subotnik_2,Gonzalez_3}
 or laser-induced transitions \cite{FISH,Mitric_1,Mitric_2,Mitric_3,SHARC} were calculated with the surface-hopping methodology.
 Spin-orbit and dipole couplings have been recently treated simultaneously with surface-hopping method \cite{SHARC,Thiel,Gonzalez_1,Gonzalez_2}.
 In every case the hopping probability is calculated from the numerical solution
 of the time-dependent Schr\"odinger equation.
 To our knowledge, there has been no surface-hopping study so far on radiative association,
 or on any kind of relaxation process that contains spontaneous emission.

 To test our method we calculate the cross sections of radiative association and
 radiative quenching for the following two reactions:

  \begin{align}
   H\left(^{2}S\right)+F\left(^{2}P\right) \rightarrow HF\left(A^{1}\Pi\right)\rightarrow \nonumber \\
    \rightarrow HF\left(A^{1}\Sigma^{+}\right)+\hbar\omega  \label{eq:1a} \\ 
   \rightarrow H\left(^{2}S\right)+F\left(^{2}P\right)+\hbar\omega \label{eq:1b}
  \end{align}

  \begin{align}
   C\left(^{3}P\right)+O\left(^{3}P\right) \rightarrow CO\left(A^{1}\Pi\right)\rightarrow \nonumber \\
    \rightarrow CO\left(A^{1}\Sigma^{+}\right)+\hbar\omega  \label{eq:2a} \\
   \rightarrow C\left(^{3}P\right)+O\left(^{3}P\right)+\hbar\omega \label{eq:2b}
  \end{align}

\noindent
 The results of our method are compared to the cross sections calculated with
 conventional semiclassical and quantum mechanical perturbation theory.
 Furthermore, the surface-hopping method allows us to calculate
 the semiclassical quantum state distributions of the stabilized diatom.
 The quantum state resolved emission spectra and the comparisons with
 the spectra obtained with quantum mechanical perturbation
 theory are also presented in section \ref{sec:Res}.

\section{\label{sec:theo} Theory}

\begin{figure}
   \includegraphics[width=8.5cm,angle=0]{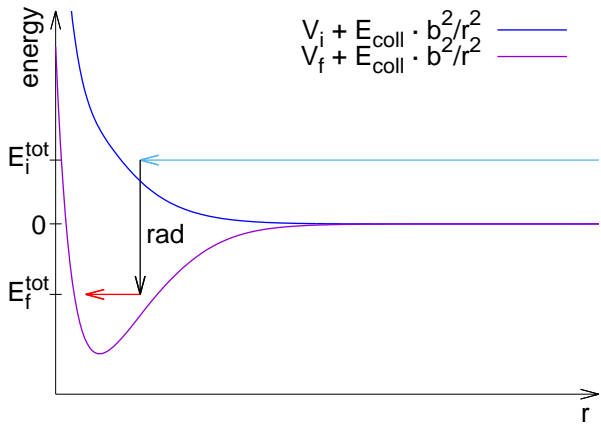}
   \caption{Schematic representation of the surface-hopping radiative association process in a diatomic system.}
   \label{fig:F0}
 \end{figure}

  Time-dependent perturbation theory provides a very useful approximation for the description of light-matter interaction.
  This brilliant tool of the time-dependent quantum mechanics is known as Fermi's golden rule. In fact, the assumptions
  of Fermi's golden rule are fulfilled in most of the problems when light absorption or emission considered \cite{Davydov,SchatzRatner}.
  It is worth to try to use it in the surface-hopping methodology to avoid the expensive numerical solution
  of the time-dependent Schr\"odinger equation in each time step. According to Fermi's golden rule \cite{Davydov,SchatzRatner}
  the probability of emission from a certain initial state to all possible
  final states with an energy is lower than $V_{i}$ is

  \begin{equation}\label{eq:3}
    P_{em}^{i\rightarrow f}\left(r\right)=\sum_{V_{f}<V_{i}}\frac{4\omega_{if}^{3}\left(r\right)}{3\hbar c^{3}}\tau\left(N_{ph}+1\right)\left|\boldsymbol{\mu}_{if}\left(r\right)\right|^{2}
  \end{equation}

  \noindent
  where the frequency of the radiation $\omega_{if}=\frac{V_{i}-V_{f}}{\hbar}$ is defined by the energy of the final and initial state, $N_{ph}$
  is the number of photons in the radiation field, $\boldsymbol{\mu}_{if}$
  is the transition dipole vector between the initial and the final electronic state, $\tau$
  is the contact-time of the interaction. Eq. (\ref{eq:3}) is valid for both stimulated and spontaneous emission,
  the latter one if there is no photon in the radiation field ($N_{ph}=0$).
  In the remainder of the paper we consider only a two-state system.
  In the surface-hopping methodology Eq. (\ref{eq:3}) is used in every time-step
  to calculate the hopping probability during the motion of the nuclei

  \begin{equation}\label{eq:4}
    P_{em}^{i\rightarrow f}\left(t\right)=\frac{4\omega_{if}^{3}\left(t\right)}{3\hbar c^{3}}{\Delta t}\left(N_{ph}+1\right)\left|\boldsymbol{\mu}_{if}\left(t\right)\right|^{2}=    A^{i\rightarrow f}(t)\Delta t
  \end{equation}

 \noindent
  where $A^{i\rightarrow f}$ is the rate of the emission.
  In this case the contact-time is the time-step of the integration of the nuclear-motion and the
  frequency of the emitted photon $\omega_{if}(t)=\frac{\max(0,V_{i}(t)-V_{f}(t))}{\hbar}$
  is the difference of the two potential surfaces and $\boldsymbol{\mu}_{if}\left(t\right)$
  is the transition dipole at the given geometry.
  With the frequency definition we implicitly assume that the classical Franck-Condon principle \cite{Franck-Condon} 
  is valid for the electronic transition,
  and the maximum function is introduced to ensure the transition from an upper to a lower state.
  A uniform random number $0\leq\xi\leq1$ is generated in each time step,
  and the hopping from the initial to the final state is performed if $P_{em}^{i\rightarrow f}(t)>\xi$.
  If we study collisions with radiative association, where there are bound states on the bottom potential energy surface,
  then we need to refine the model. It is possible that the energy of the emitted photon ($\hbar\omega_{if}$)
  is not high enough for the association of fragments, because after hopping the energy content
  of the system can be higher than the dissociation energy on the bottom potential energy surface.
  Thus we need to distinguish the two channels after a hopping:

  \begin{equation} \label{eq:5}
   \textrm{if}\left\{ \begin{array}{c}
   \hbar\omega>E_{i}^{tot}\;\textrm{and}\;V_{f}^{eff}\left(t\right)<0\;\;\textrm{then}\;\textrm{associaton}\\
     \\
   \hbar\omega<E_{i}^{tot}\;\;\textrm{or}\;V_{f}^{eff}\left(t\right)>0\;\;\textrm{then}\;\textrm{quenching}\\
   \end{array}\right.
  \end{equation}

 \noindent
   i) radiative association:
   if the initial total energy (total energy of the system before collision on the upper potential energy surface)
   is smaller than the energy of the photon and the bottom effective potential surface has bound states;
   ii) radiative quenching:
   the fragments dissociate after the hopping. 
   The former process is illustrated schematically in Fig.~\ref{fig:F0}.
   $V_{f}^{eff}\left(t\right)$ in Eq. (\ref{eq:5}) is the effective potential energy surface of 
   the final state when a hopping occures:

  \begin{equation}
   V_{f}^{eff}\left(t\right)=V_{f}\left(t\right)+E_{coll}\frac{b^{2}}{r^{2}\left(t\right)}
  \end{equation}
   where $E_{coll}$ is the collision energy, $b$ is the impact paramter of the collision and
   $r$ is the distance between atoms.

  After generation of $N_{tot}$
  number of trajectories at a given $\left(E_{coll},b\right)$
  pair the probability of the stimulated emission (the opacity function) can be obtained

 \begin{equation} \label{eq:6}
   P_{rad}^{st}\left(E_{coll},b\right)\approx\frac{N_{hop}\left(E_{coll},b\right)}{N_{tot}\left(E_{coll},b\right)}
 \end{equation}

 \noindent
  where $N_{hop}\left(E_{coll},b\right)$
  is the number of trajectories where hopping was detected.
  If one would like to simulate spontaneous emission then $N_{ph}=0$
  should be considered. In this case, however, the probability of a transitions would
  be very small and $10^{8}-10^{10}$ trajectories might be needed to estimate $P_{rad}^{sp}$
  in a general molecular system. We can bypass this obstacle if transitions with stimulated emission
  is calculated even when the aim is spontaneous emission. In every time step we calculate the hopping
  probability with a certain photon number. To get the probability of spontaneous
  emission we need to scale down the obtained probability of stimulated emission with the applied photon number

 \begin{equation} \label{eq:7}
   P_{rad}^{sp}\left(E_{coll},b\right)\approx\left(\frac{1}{N_{ph}+1}\right)\frac{N_{hop}\left(E_{coll},b\right)}{N_{tot}\left(E_{coll},b\right)}
 \end{equation}

 \noindent
  The trick with the adjustable photon number is profitable in the simulation of spontaneous
  emission. The required trajectory number in the estimation of $P_{rad}^{sp}$
  (and at the same time also the computational time) can be optimized by variation of $N_{ph}$.
   From this point the cross section of the radiative processes

 \begin{equation} \label{eq:8}
   \sigma\left(E_{coll}\right)=2\pi f_{stat}\intop_{0}^{\infty}bP_{rad}\left(E_{coll},b\right)db
 \end{equation}

 \noindent
  can be calculated, where $f_{stat}$ is the statistical weight factor
  for the initial molecular state of the approaching fragments.

  It is easy to prove that the surface-hopping method is equivalent with the conventional semiclassical
  method used in former studies for diatomics \cite{Dalgarno}. Eq. (\ref{eq:4}) is considered only for one time-slice of the whole collision process,
  when Eq. (\ref{eq:4}) is integrated over the time of the collision then we get $P_{rad}$.
  If the time-variable is changed to distance in the integral, the Jacobi-determinant
  of the variable-transformation will be the reciprocal of the velocity. The result of the
  variable transformation is the well known semiclassical formula

  \begin{eqnarray} \label{eq:9}
   P_{rad}  =  \intop A^{i\rightarrow f}(t)dt=\intop\frac{A^{i\rightarrow f}(r)}{v}dr= \nonumber \\
     =  \sqrt{\frac{\mu}{2}}\intop\frac{A^{i\rightarrow f}(r)}{\sqrt{E_{coll}-E_{coll}\frac{b^{2}}{r^{2}}-V_{i}(r)}}dr
  \end{eqnarray}

  \noindent
  as pointed out by Kramers and ter Haar seven decades ago \cite{Kramers}.
  This allows us to consider the surface-hopping method as the extension of the former semiclassical
  method in arbitrary dimension.

  Furthermore, contrary to the former semiclassical formula
  the surface-hopping methodology provides real dynamical
  coupling between molecular states and the radiation field:
  after hopping we can follow the fate of the trajectory. That means we can characterize
  the ro-vibrational states of the formed molecules after each reactive hopping.
  The WKB quantization can be applied for the stabilized diatom
  (as a rotating anharmonic oscillator) to obtain the semiclassical vibrational quantum number \cite{Semiquant} :

  \begin{equation} \label{eq:10}
   v'=\frac{\sqrt{2\mu}}{\hbar \pi } \intop_{r_{min}}^{r_{max}}\left(E_{f}^{tot}-\frac{L^{2}}{2\mu r^{2}}-V_{f}(r)\right)^{\frac{1}{2}}dr -\frac{1}{2}
  \end{equation}
  
  \noindent
  where $\mu$ is the reduces mass and $L$ is the angular momentum of the diatom, $r_{min}$ and $r_{max}$ are the turningpoints,
  and $E_{f}^{tot}$ is the total energy content of the system after a reactive hopping.
  The quantum state resolved cross sections or emission spectra can be calculated
  after the semiclassical quantization. That means the surface-hopping method can
  provide the most detailed levels of the characterization of radiative association
  using semiclassical mechanics.

  The modeling of radiative association with surface-hopping methodology has an additional
  advantage compared with the conventional surface-hopping method used in the simulation
  of nonadibatic dynamics: the scaling of momenta is not needed because the photon takes away
  energy from the molecular system, equal to the change in potential energy.
  This is again due to the application of the classical Franck-Condon principle.
  The former surface-hopping methods (including nonadiabatic and spin-orbit couplings) extended
  with our model could provide an efficient and complete description of the coupled photophysical
  processes in molecular collisions or in molecular systems after an electronic excitation.

\section{\label{sec:comp}Computational details}

  Hamilton's equations of motion were used in Cartesian coordinates for the calculation of trajectories.
  A fourth order symplectic integrator \cite{Sympfour} was employed for integration with a step size of 0.05 fs.
  The initial separation between the atoms was fixed at 18 bohr,
  which guaranteed a negligible initial interaction between the colliding atoms.
  The hopping probability in each time step was calculated according to Eq. (\ref{eq:4}) and
  the cross sections were calculated according to the Eq. (\ref{eq:8}), where the integral over impact parameter
  was evaluated with Simpson's 1/3 rule. The value of $b_{max}$ (the upper bound) in the integral in Eq. (\ref{eq:8}) was adjusted at each collision energy to account for all hopping events.
  $N_{tot}$=1000 trajectories with adjusted photon number ($N_{ph}$) were used to estimate the probability
  of radiation ($P_{rad}(E_{coll},b)$) at each ($E_{coll},b$). It should be noted that the photon number used in
  computation of the stimulated emission, Eq. (\ref{eq:4}) should be small enough to avoid saturation 
  of the hopping, which in turn results in an underestimation of
  $P_{rad}(E_{coll},b)$ through the division by $N_{ph}+1$ in Eq.(\ref{eq:7}).
  As a rule of thumb we suggest that the photon number should be set to
  produce less hopping events than 20-30\% of $N_{tot}$. 
  In the calculation of state-resolved spectra we ran a total of $10^{6}$ trajectories to get satisfying
  statistics. 
  We used a bin size of $\Delta E_{photon}=0.1$ eV for the calculation of the spectral densities.
  The rotational quantum number of the products, $j'$, was calculated from the conventional
  semiclassial qunatization formula $L=\hbar\sqrt{j'(j'+1)}$.
  The quasiclassical vibrational quantum number was calculated according to Eq. (\ref{eq:10}) where
  the integration  over the interatomic distance was evaluated with trapezoidal rule. The trajectories were propagated
  for more 500 fs after a reactive hopping to determine the turning-points used in the calculation
  of the action-integral. Eq. (\ref{eq:10}) gives continuous quantum numbers. 
  To obtain discrete quantum numbers we used the standard binning technique:
  each trajectory contributes to $P_{rad}(E_{coll},b)$ with equal weight and
  the quantum numbers are rounded to the nearest integer.
  The details of the conventional semiclassical and quantum mechanical
  perturbation theory used for comparison can be found in Refs.~\onlinecite{HF_paper,CO_paper1,CO_paper2}.

\section{\label{sec:Res}Results}

\subsection{Cross sections}
\begin{figure}
   \includegraphics[width=8.5cm,angle=0]{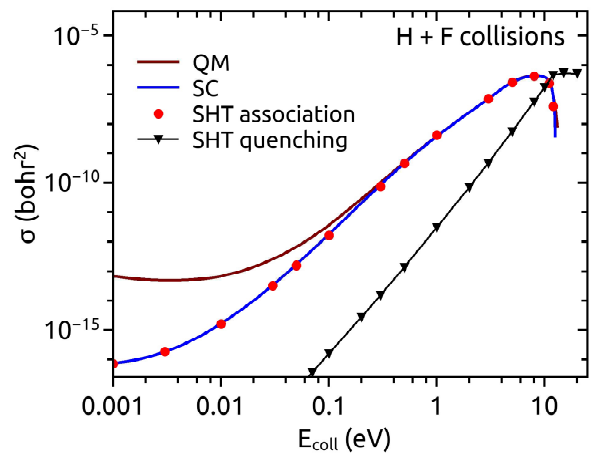}
   \caption{Cross sections for the radiative association and quenching of H + F collisions
            according to the reaction (\ref{eq:1a}) and (\ref{eq:1b}), respectively.}
   \label{fig:F1}
 \end{figure}

 First consider the HF test system. Since it contains one low mass atom
 a significant tunneling effect is expected at low energies. Moreover the potential energy
 surface of the excited state is almost entirely repulsive \cite{HF_paper}.
 Fig.~\ref{fig:F1} shows the cross sections of radiative association through Eq.(\ref{eq:8})
 computed with three different methods: quantum mechanical perturbation theory (QM),
 conventional semiclassical method (SC) and our surface-hopping trajectory
 method (SHT). 

 The SHT cross sections perfectly match the curve
 obtained with the conventional semiclassical method in the whole collision energy range.
 Resonance structure can not be observed on QM curve because the upper potential
 energy surface is repulsive. Due to this feature of the HF system 
 the dominant quantum mechanical effect is the tunneling into the
 repulsive region of the excited potential surface.
 The semiclassical methods show good agreement with the QM cross sections
 when $E_{coll} > 0.2$ eV. The semiclassical methods underestimate the QM results
 in the low energy domains. This shows the importance of tunneling below $E_{coll}=0.2$ eV. 
 Fig.~\ref{fig:F1} also shows the SHT cross sections for the quenching channel
 where electronic transition was detected during the collisions but the HF
 molecules were not able to stabilize due to the remaining high energy content in the system.

 The cross section of the quenching channel is smaller almost in the whole energy range
 with 2-3 orders of magnitude than that of association. Over $E_{coll}=10$ eV
 the quenching process dominates, and at the same time the probability of
 the association is suddenly diminishing. Above $E_{coll}=10$ eV the probability of
 quenching is an essentially constant function of $E_{coll}$ and above $E_{coll}=$ 11-13 eV
 the ionization and electronic excitation of the atoms may play a big role \cite{NIST_ASD}.

\begin{figure}
    \includegraphics[width=8.5cm,angle=0]{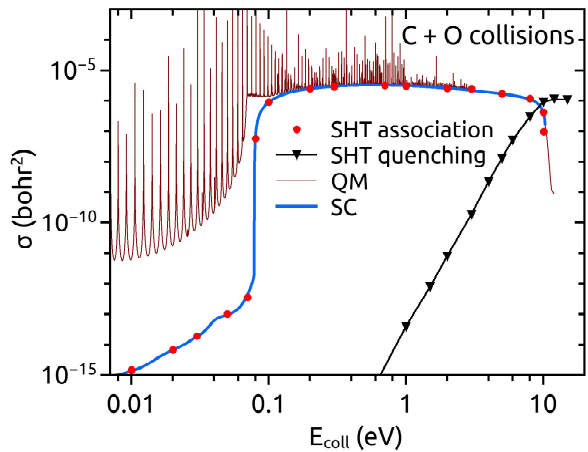}
    \caption{Cross sections for the radiative association and quenching of C + O collisions
             according to the reaction (\ref{eq:2a}) and (\ref{eq:2b}), respectively.}
   \label{fig:F2}
 \end{figure}

 The second test system, CO, is somewhat different from HF.
 The excited potential energy surface is attractive and has many bound states \cite{CO_paper1,CO_paper2}.
 In this case shape resonances, due to the tunneling into the quasibound states, are expected.
 The SHT method shows again perfect matching with the conventional semiclassical method (Fig.~\ref{fig:F2}).
 At high energies good agreement can be found between the QM and the semiclassical methods.
 The resonance contribution is quite relevant under $E_{coll}=1.0$ eV, but the semiclassical methods
 reproduce the QM base line at energies down to about 0.1 eV. Below $E_{coll}=0.1$ eV, however,
 the SC and SHT cross sections are smaller than that from QM with a factor of $10^{3}-10^{4}$.
 The behavior of the quenching channel is similar to that of the HF system. Below $E_{coll}=9.0$ eV
 the radiative quenching curve shows a falloff shape, at $E_{coll}=9.0$ eV it is equally probable as
 the association channel. The edges of atomic 
 ionization and excitation channels start also about at this energy value \cite{NIST_ASD}.
 The ionization and excitation could be the dominant processes over 10-13 eV.

\subsection{Emission spectra}

\begin{figure}
  \includegraphics[width=8.5cm,angle=0]{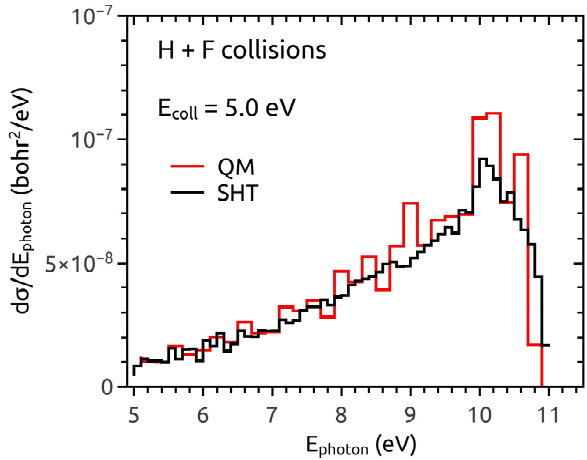}
  \caption{Spectral density for radiative association of HF through reaction (\ref{eq:1a})
          The SHT result is computed as explained in section~\ref{sec:comp}.
          The QM result is computed from the QM data presented in Fig.~\ref{fig:F5}
          by adding the cross sections in photon energy intervals of $\Delta E_{photon}=0.2$ eV
          and then divided the sum by $\Delta E_{photon}$.}
  \label{fig:F3}
 \end{figure}

 The radiative association process can also be characterized by emission spectra.
 Fig.~\ref{fig:F3} shows the collision induced emission spectrum for the HF system at $E_{coll}=5.0$ eV.
 The SHT emission spectrum agrees well with that is calculated by quantum mechanical perturbation theory.
 It means that the SHT method is able to reproduce also the energy distribution of photons,
 not only the cross sections as a function of $E_{coll}$.
 Some differences can be seen in the high energy domains where the QM
 spectrum has a fine structure that the SHT method misses.

\begin{figure}
  \includegraphics[width=8.5cm,angle=0]{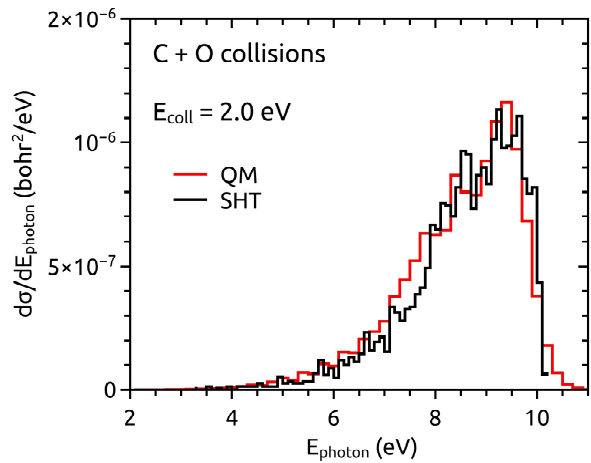}
  \caption{As in Fig. ~\ref{fig:F3} but for CO formation through reaction (\ref{eq:2a}).
          The QM result is computed from the QM data in Fig.~\ref{fig:F6}.}
  \label{fig:F4}
 \end{figure}

 Fig.~\ref{fig:F4} shows the spectral densities of radiative association for the CO system. 
 The SHT method performs well also for this reaction. The SHT and QM
 spectra match in the whole energy range. 

 \begin{figure}
  \includegraphics[width=7.5cm,angle=0]{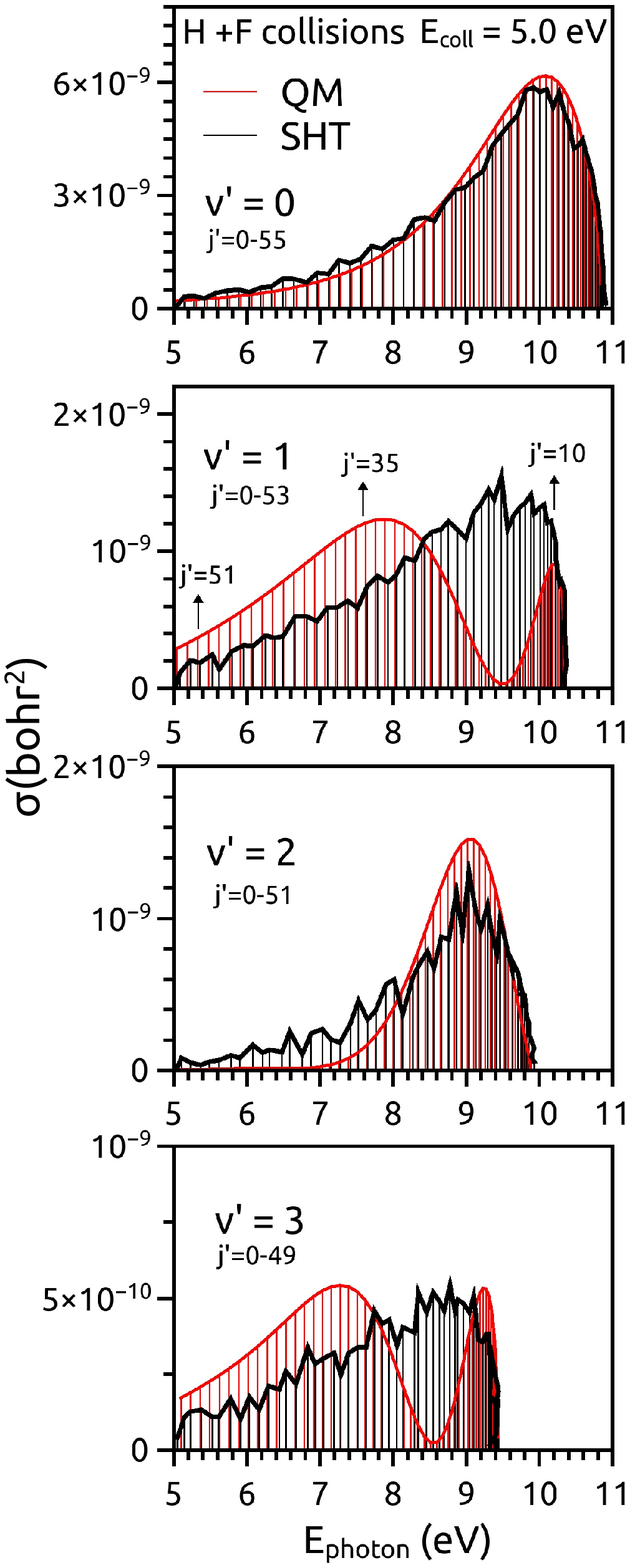}
  \caption{Rovibrationally state resolved emission spectra for radiative association of HF through reaction (\ref{eq:1a}).
          The sticks correspond to the ($v',j'$) states.}
  \label{fig:F5}
 \end{figure}

 \begin{figure}
  \includegraphics[width=7.5cm,angle=0]{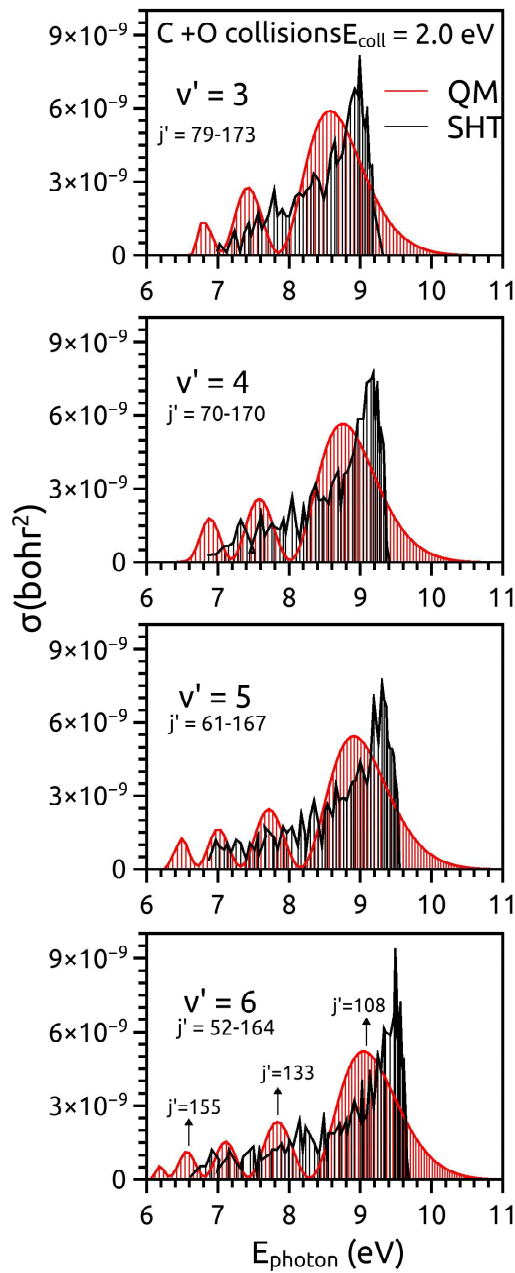}
  \caption{Rovibrationally state resolved emission spectra for radiative association of CO through reaction (\ref{eq:2a}).
          The sticks correspond to the ($v',j'$) states.}
  \label{fig:F6}
 \end{figure}

 The surface-hopping method has another big advantage: it allows us to
 calculate the quantum state resolved spectra or cross sections with
 semiclassical quantization. 
 Fig.~\ref{fig:F5} shows the comparisons of the
 rovibrationally state resolved emission spectra for reaction (\ref{eq:1a}) obtained with SHT and QM method.
 The envelope of SHT and QM spectra for $v'=0$ match well. The SHT method slightly
 underestimates the QM curve only in the high energy region.
 The QM spectra that corresponds to the higher vibrational quantum states
 show more complex character. 
 The SHT method can not reproduce the detailed structure of the QM
 spectra, but overall the spectral distributions are similar.
 If we add the sticks, each corresponding to one rotational quantum
 number, then the obtained vibrationally resolved SHT cross sections do
 agree well with the cross sections of the QM method.
 Fig.~\ref{fig:F6} shows the state resolved emission spectra at $E_{coll}=$ 2 eV for reaction (\ref{eq:2a}).
 The four vibrational states were plotted that have the biggest contribution 
 to the total/unresolved spectrum.
 These QM spectra have more complex character than in the H + F collisions.
 In this case the envelopes of SHT and QM curves agree not so well than in reaction (\ref{eq:1a}),
 but the SHT curves follow the average shape of the QM curves in every case.
 Most of the products are born with huge rotational and with moderate vibrational
 quantum number. 

 These results indicate that the surface-hopping method
 provides reasonable result even for the most detailed quantum state distributions
 that characterize the collision induced emission.

\section{Conclusion}

 In this work we have presented a semiclassical method based on surface-hopping methodology
 for the modeling of radiative association with electronic transitions.
 In the surface-hopping formalism we use Fermi's golden rule to
 calculate the probability of hopping during the propagation of nuclei.
 Due to Fermi's golden rule we can avoid to solve numerically the time-dependent
 Schr\"odinger equation in each time step. It can be proven that our method, 
 when implemented for two colliding atoms, is equivalent to the established
 semiclassical formula for diatomic systems.
 But, contrary to the latter, the surface-hopping methodology 
 allows us to extend the study of radiative association
 with electronic transitions to arbitrary polyatomic systems, and also to follow the
 fate of trajectories after a photon emission. Due to this property of the method
 we can characterize the rovibrational states of the stabilized molecules.
 To test our method we have calculated the cross sections of radiative association and
 radiative quenching for two diatomic reactions. The surface-hopping
 method agrees with the conventional semiclassical formula. This
 was expected due to the equivalence of the two methods.
 Moreover the obtained total and the quantum state resolved spectra
 are compared to the results of a quantum mechanical perturbation theory.
 These comparisons imply that the surface-hopping method extended with
 Fermi's golden rule can provide reasonable result even for the most detailed quantum state
 distributions for the characterization of radiative association.
 In addition our surface-hopping model is not only applicable for
 the description of radiative association but it can be used for
 a numerically inexpensive semiclassical characterization of any molecular process
 where spontaneous emission occurs.

\begin{acknowledgments}
 Support from the Kempe Foundation is gratefully acknowledged.
\end{acknowledgments}

\nocite{*}

\bibliography{surface-hopping}

\providecommand{\noopsort}[1]{}\providecommand{\singleletter}[1]{#1}%
\begin{thebibliography}{62}%
\makeatletter
\providecommand \@ifxundefined [1]{%
 \@ifx{#1\undefined}
}%
\providecommand \@ifnum [1]{%
 \ifnum #1\expandafter \@firstoftwo
 \else \expandafter \@secondoftwo
 \fi
}%
\providecommand \@ifx [1]{%
 \ifx #1\expandafter \@firstoftwo
 \else \expandafter \@secondoftwo
 \fi
}%
\providecommand \natexlab [1]{#1}%
\providecommand \enquote  [1]{``#1''}%
\providecommand \bibnamefont  [1]{#1}%
\providecommand \bibfnamefont [1]{#1}%
\providecommand \citenamefont [1]{#1}%
\providecommand \href@noop [0]{\@secondoftwo}%
\providecommand \href [0]{\begingroup \@sanitize@url \@href}%
\providecommand \@href[1]{\@@startlink{#1}\@@href}%
\providecommand \@@href[1]{\endgroup#1\@@endlink}%
\providecommand \@sanitize@url [0]{\catcode `\\12\catcode `\$12\catcode
  `\&12\catcode `\#12\catcode `\^12\catcode `\_12\catcode `\%12\relax}%
\providecommand \@@startlink[1]{}%
\providecommand \@@endlink[0]{}%
\providecommand \url  [0]{\begingroup\@sanitize@url \@url }%
\providecommand \@url [1]{\endgroup\@href {#1}{\urlprefix }}%
\providecommand \urlprefix  [0]{URL }%
\providecommand \Eprint [0]{\href }%
\providecommand \doibase [0]{http://dx.doi.org/}%
\providecommand \selectlanguage [0]{\@gobble}%
\providecommand \bibinfo  [0]{\@secondoftwo}%
\providecommand \bibfield  [0]{\@secondoftwo}%
\providecommand \translation [1]{[#1]}%
\providecommand \BibitemOpen [0]{}%
\providecommand \bibitemStop [0]{}%
\providecommand \bibitemNoStop [0]{.\EOS\space}%
\providecommand \EOS [0]{\spacefactor3000\relax}%
\providecommand \BibitemShut  [1]{\csname bibitem#1\endcsname}%
\let\auto@bib@innerbib\@empty
\bibitem [{\citenamefont {Babb}\ and\ \citenamefont
  {Kirby}(1998)}]{Babb_Kirby_98}%
  \BibitemOpen
  \bibfield  {author} {\bibinfo {author} {\bibfnamefont {J.~F.}\ \bibnamefont
  {Babb}}\ and\ \bibinfo {author} {\bibfnamefont {K.~P.}\ \bibnamefont
  {Kirby}},\ }in\ \href@noop {} {\emph {\bibinfo {booktitle} {The Molecular
  Astrophysics of Stars and Galaxies}}},\ \bibinfo {editor} {edited by\
  \bibinfo {editor} {\bibfnamefont {T.~W.}\ \bibnamefont {Hartquist}}\ and\
  \bibinfo {editor} {\bibfnamefont {D.~A.}\ \bibnamefont {Williams}}}\
  (\bibinfo  {publisher} {Clarendon {P}ress},\ \bibinfo {address} {Oxford},\
  \bibinfo {year} {1998})\ p.~\bibinfo {pages} {11}\BibitemShut {NoStop}%
\bibitem [{\citenamefont {Gerlich}\ and\ \citenamefont
  {Horning}(1992)}]{Gerlich_92_cr_92_1509}%
  \BibitemOpen
  \bibfield  {author} {\bibinfo {author} {\bibfnamefont {D.}~\bibnamefont
  {Gerlich}}\ and\ \bibinfo {author} {\bibfnamefont {S.}~\bibnamefont
  {Horning}},\ }\href@noop {} {\bibfield  {journal} {\bibinfo  {journal}
  {Chem.\ Rev.}\ }\textbf {\bibinfo {volume} {92}},\ \bibinfo {pages} {1509}
  (\bibinfo {year} {1992})}\BibitemShut {NoStop}%
\bibitem [{\citenamefont {Nyman}, \citenamefont {Gustafsson},\ and\
  \citenamefont {Antipov}(2015)}]{Rev_theo}%
  \BibitemOpen
  \bibfield  {author} {\bibinfo {author} {\bibfnamefont {G.}~\bibnamefont
  {Nyman}}, \bibinfo {author} {\bibfnamefont {M.}~\bibnamefont {Gustafsson}}, \
  and\ \bibinfo {author} {\bibfnamefont {S.~V.}\ \bibnamefont {Antipov}},\
  }\href@noop {} {\bibfield  {journal} {\bibinfo  {journal} {Int.\ Rev.\ Phys.\
  Chem.}\ }\textbf {\bibinfo {volume} {34}},\ \bibinfo {pages} {385} (\bibinfo
  {year} {2015})}\BibitemShut {NoStop}%
\bibitem [{\citenamefont {Czak\'o}, \citenamefont {Szab\'o},\ and\
  \citenamefont {Telkes}(2014)}]{Czako_1}%
  \BibitemOpen
  \bibfield  {author} {\bibinfo {author} {\bibfnamefont {G.}~\bibnamefont
  {Czak\'o}}, \bibinfo {author} {\bibfnamefont {I.}~\bibnamefont {Szab\'o}}, \
  and\ \bibinfo {author} {\bibfnamefont {H.}~\bibnamefont {Telkes}},\
  }\href@noop {} {\bibfield  {journal} {\bibinfo  {journal} {J.\ Phys.\ Chem.\
  A.}\ }\textbf {\bibinfo {volume} {118}},\ \bibinfo {pages} {646} (\bibinfo
  {year} {2014})}\BibitemShut {NoStop}%
\bibitem [{\citenamefont {Czak\'o}\ and\ \citenamefont
  {Bowman}(2014)}]{Czako_2}%
  \BibitemOpen
  \bibfield  {author} {\bibinfo {author} {\bibfnamefont {G.}~\bibnamefont
  {Czak\'o}}\ and\ \bibinfo {author} {\bibfnamefont {J.~M.}\ \bibnamefont
  {Bowman}},\ }\href@noop {} {\bibfield  {journal} {\bibinfo  {journal} {J.\
  Phys.\ Chem.\ A.}\ }\textbf {\bibinfo {volume} {118}},\ \bibinfo {pages}
  {2839} (\bibinfo {year} {2014})}\BibitemShut {NoStop}%
\bibitem [{\citenamefont {Yurchenko}\ \emph {et~al.}(2009)\citenamefont
  {Yurchenko}, \citenamefont {Yachmenev}, \citenamefont {Thiel}, \citenamefont
  {Baum}, \citenamefont {Giesen}, \citenamefont {Melnikov},\ and\ \citenamefont
  {Jensen}}]{Jensen}%
  \BibitemOpen
  \bibfield  {author} {\bibinfo {author} {\bibfnamefont {S.~N.}\ \bibnamefont
  {Yurchenko}}, \bibinfo {author} {\bibfnamefont {A.}~\bibnamefont
  {Yachmenev}}, \bibinfo {author} {\bibfnamefont {W.}~\bibnamefont {Thiel}},
  \bibinfo {author} {\bibfnamefont {O.}~\bibnamefont {Baum}}, \bibinfo {author}
  {\bibfnamefont {T.~F.}\ \bibnamefont {Giesen}}, \bibinfo {author}
  {\bibfnamefont {V.~V.}\ \bibnamefont {Melnikov}}, \ and\ \bibinfo {author}
  {\bibfnamefont {P.}~\bibnamefont {Jensen}},\ }\href@noop {} {\bibfield
  {journal} {\bibinfo  {journal} {J.\ Mol.\ Spec.}\ }\textbf {\bibinfo {volume}
  {257}},\ \bibinfo {pages} {57} (\bibinfo {year} {2009})}\BibitemShut
  {NoStop}%
\bibitem [{\citenamefont {Zak}\ \emph {et~al.}(2016)\citenamefont {Zak},
  \citenamefont {Tennyson}, \citenamefont {Polyansky}, \citenamefont {Lodi},
  \citenamefont {Zobov}, \citenamefont {Tashkun},\ and\ \citenamefont
  {Perevalov}}]{Tennyison}%
  \BibitemOpen
  \bibfield  {author} {\bibinfo {author} {\bibfnamefont {E.}~\bibnamefont
  {Zak}}, \bibinfo {author} {\bibfnamefont {J.}~\bibnamefont {Tennyson}},
  \bibinfo {author} {\bibfnamefont {O.~L.}\ \bibnamefont {Polyansky}}, \bibinfo
  {author} {\bibfnamefont {L.}~\bibnamefont {Lodi}}, \bibinfo {author}
  {\bibfnamefont {N.~F.}\ \bibnamefont {Zobov}}, \bibinfo {author}
  {\bibfnamefont {S.~A.}\ \bibnamefont {Tashkun}}, \ and\ \bibinfo {author}
  {\bibfnamefont {V.~I.}\ \bibnamefont {Perevalov}},\ }\href@noop {} {\bibfield
   {journal} {\bibinfo  {journal} {J.\ Mol.\ Spec.}\ }\textbf {\bibinfo
  {volume} {177}},\ \bibinfo {pages} {31} (\bibinfo {year} {2016})}\BibitemShut
  {NoStop}%
\bibitem [{\citenamefont {Lique}\ \emph {et~al.}(2009)\citenamefont {Lique},
  \citenamefont {Jorfi}, \citenamefont {Honvault}, \citenamefont {Halvick},
  \citenamefont {Lin}, \citenamefont {Guo},\ and\ \citenamefont {Xie}}]{OHO}%
  \BibitemOpen
  \bibfield  {author} {\bibinfo {author} {\bibfnamefont {F.}~\bibnamefont
  {Lique}}, \bibinfo {author} {\bibfnamefont {M.}~\bibnamefont {Jorfi}},
  \bibinfo {author} {\bibfnamefont {P.}~\bibnamefont {Honvault}}, \bibinfo
  {author} {\bibfnamefont {P.}~\bibnamefont {Halvick}}, \bibinfo {author}
  {\bibfnamefont {S.~Y.}\ \bibnamefont {Lin}}, \bibinfo {author} {\bibfnamefont
  {H.}~\bibnamefont {Guo}}, \ and\ \bibinfo {author} {\bibfnamefont {D.~Q.}\
  \bibnamefont {Xie}},\ }\href@noop {} {\bibfield  {journal} {\bibinfo
  {journal} {J.\ Chem.\ Phys.}\ }\textbf {\bibinfo {volume} {131}},\ \bibinfo
  {pages} {221104} (\bibinfo {year} {2009})}\BibitemShut {NoStop}%
\bibitem [{\citenamefont {Antipov}\ \emph {et~al.}(2009)\citenamefont
  {Antipov}, \citenamefont {Sj{\"o}lander}, \citenamefont {Nyman},\ and\
  \citenamefont {Gustafsson}}]{Antipov_09_jcp}%
  \BibitemOpen
  \bibfield  {author} {\bibinfo {author} {\bibfnamefont {S.~V.}\ \bibnamefont
  {Antipov}}, \bibinfo {author} {\bibfnamefont {T.}~\bibnamefont
  {Sj{\"o}lander}}, \bibinfo {author} {\bibfnamefont {G.}~\bibnamefont
  {Nyman}}, \ and\ \bibinfo {author} {\bibfnamefont {M.}~\bibnamefont
  {Gustafsson}},\ }\href@noop {} {\bibfield  {journal} {\bibinfo  {journal}
  {J.\ Chem.\ Phys.}\ }\textbf {\bibinfo {volume} {131}},\ \bibinfo {pages}
  {074302} (\bibinfo {year} {2009})}\BibitemShut {NoStop}%
\bibitem [{\citenamefont {Antipov}, \citenamefont {Gustafsson},\ and\
  \citenamefont {Nyman}(2011)}]{Antipov_11_jcp}%
  \BibitemOpen
  \bibfield  {author} {\bibinfo {author} {\bibfnamefont {S.~V.}\ \bibnamefont
  {Antipov}}, \bibinfo {author} {\bibfnamefont {M.}~\bibnamefont {Gustafsson}},
  \ and\ \bibinfo {author} {\bibfnamefont {G.}~\bibnamefont {Nyman}},\
  }\href@noop {} {\bibfield  {journal} {\bibinfo  {journal} {J.\ Chem.\ Phys.}\
  }\textbf {\bibinfo {volume} {135}},\ \bibinfo {pages} {184302} (\bibinfo
  {year} {2011})}\BibitemShut {NoStop}%
\bibitem [{\citenamefont {Barinovs}\ and\ \citenamefont {van
  Hemert}(2005)}]{Barinovs_05_ApJ_636_923}%
  \BibitemOpen
  \bibfield  {author} {\bibinfo {author} {\bibfnamefont {{\v{G}}.}~\bibnamefont
  {Barinovs}}\ and\ \bibinfo {author} {\bibfnamefont {M.~C.}\ \bibnamefont {van
  Hemert}},\ }\href@noop {} {\bibfield  {journal} {\bibinfo  {journal}
  {Astrophys.\ J.}\ }\textbf {\bibinfo {volume} {636}},\ \bibinfo {pages} {923}
  (\bibinfo {year} {2005})}\BibitemShut {NoStop}%
\bibitem [{\citenamefont {Bennett}\ \emph {et~al.}(2003)\citenamefont
  {Bennett}, \citenamefont {Dickinson}, \citenamefont {Leininger},\ and\
  \citenamefont {Gad{\'e}a}}]{Bennett_03_MonNotRAstronSoc_341_361}%
  \BibitemOpen
  \bibfield  {author} {\bibinfo {author} {\bibfnamefont {O.~J.}\ \bibnamefont
  {Bennett}}, \bibinfo {author} {\bibfnamefont {A.~S.}\ \bibnamefont
  {Dickinson}}, \bibinfo {author} {\bibfnamefont {T.}~\bibnamefont
  {Leininger}}, \ and\ \bibinfo {author} {\bibfnamefont {F.~X.}\ \bibnamefont
  {Gad{\'e}a}},\ }\href@noop {} {\bibfield  {journal} {\bibinfo  {journal}
  {Mon.\ Not.\ R.\ Astron.\ Soc.}\ }\textbf {\bibinfo {volume} {341}},\
  \bibinfo {pages} {361} (\bibinfo {year} {2003})},\ \bibinfo {note} {erratum:
  Mon.\ Not.\ R.\ Astron.\ Soc. {\bf 384} 1743 (2008)}\BibitemShut {NoStop}%
\bibitem [{\citenamefont {Dickinson}(2005)}]{Dickinson_05_JPhysB_38_4329}%
  \BibitemOpen
  \bibfield  {author} {\bibinfo {author} {\bibfnamefont {A.~S.}\ \bibnamefont
  {Dickinson}},\ }\href@noop {} {\bibfield  {journal} {\bibinfo  {journal} {J.\
  Phys.\ B}\ }\textbf {\bibinfo {volume} {38}},\ \bibinfo {pages} {4329}
  (\bibinfo {year} {2005})},\ \bibinfo {note} {corrigendum: J.\ Phys.\ B {\bf
  41} 049801 (2008)}\BibitemShut {NoStop}%
\bibitem [{\citenamefont {Franz}, \citenamefont {Gustafsson},\ and\
  \citenamefont {Nyman}(2011)}]{CO_paper1}%
  \BibitemOpen
  \bibfield  {author} {\bibinfo {author} {\bibfnamefont {J.}~\bibnamefont
  {Franz}}, \bibinfo {author} {\bibfnamefont {M.}~\bibnamefont {Gustafsson}}, \
  and\ \bibinfo {author} {\bibfnamefont {G.}~\bibnamefont {Nyman}},\
  }\href@noop {} {\bibfield  {journal} {\bibinfo  {journal} {Mon.\ Not.\ R.\
  Astron.\ Soc.}\ }\textbf {\bibinfo {volume} {414}},\ \bibinfo {pages} {3547}
  (\bibinfo {year} {2011})}\BibitemShut {NoStop}%
\bibitem [{\citenamefont {Gustafsson}, \citenamefont {Monge-Palacios},\ and\
  \citenamefont {Nyman}(2014)}]{HF_paper}%
  \BibitemOpen
  \bibfield  {author} {\bibinfo {author} {\bibfnamefont {M.}~\bibnamefont
  {Gustafsson}}, \bibinfo {author} {\bibfnamefont {M.}~\bibnamefont
  {Monge-Palacios}}, \ and\ \bibinfo {author} {\bibfnamefont {G.}~\bibnamefont
  {Nyman}},\ }\href@noop {} {\bibfield  {journal} {\bibinfo  {journal} {J.\
  Chem.\ Phys.}\ }\textbf {\bibinfo {volume} {140}},\ \bibinfo {pages} {184301}
  (\bibinfo {year} {2014})}\BibitemShut {NoStop}%
\bibitem [{\citenamefont {Gustafsson}\ and\ \citenamefont
  {Nyman}(2015)}]{CO_paper2}%
  \BibitemOpen
  \bibfield  {author} {\bibinfo {author} {\bibfnamefont {M.}~\bibnamefont
  {Gustafsson}}\ and\ \bibinfo {author} {\bibfnamefont {G.}~\bibnamefont
  {Nyman}},\ }\href@noop {} {\bibfield  {journal} {\bibinfo  {journal} {Mon.\
  Not.\ R.\ Astron.\ Soc.}\ }\textbf {\bibinfo {volume} {448}},\ \bibinfo
  {pages} {2562} (\bibinfo {year} {2015})}\BibitemShut {NoStop}%
\bibitem [{\citenamefont {Singh}\ and\ \citenamefont
  {Andreazza}(2000)}]{Singh_00_ApJ_537_261}%
  \BibitemOpen
  \bibfield  {author} {\bibinfo {author} {\bibfnamefont {P.~D.}\ \bibnamefont
  {Singh}}\ and\ \bibinfo {author} {\bibfnamefont {C.~M.}\ \bibnamefont
  {Andreazza}},\ }\href@noop {} {\bibfield  {journal} {\bibinfo  {journal}
  {Astrophys.\ J.}\ }\textbf {\bibinfo {volume} {537}},\ \bibinfo {pages} {261}
  (\bibinfo {year} {2000})}\BibitemShut {NoStop}%
\bibitem [{\citenamefont {Mruga{\l}a}, \citenamefont {{\v{S}}pirko},\ and\
  \citenamefont {Kraemer}(2003)}]{Mrugala_2003}%
  \BibitemOpen
  \bibfield  {author} {\bibinfo {author} {\bibfnamefont {F.}~\bibnamefont
  {Mruga{\l}a}}, \bibinfo {author} {\bibfnamefont {V.}~\bibnamefont
  {{\v{S}}pirko}}, \ and\ \bibinfo {author} {\bibfnamefont {W.~P.}\
  \bibnamefont {Kraemer}},\ }\href@noop {} {\bibfield  {journal} {\bibinfo
  {journal} {J.\ Chem.\ Phys.}\ }\textbf {\bibinfo {volume} {118}},\ \bibinfo
  {pages} {10547} (\bibinfo {year} {2003})}\BibitemShut {NoStop}%
\bibitem [{\citenamefont {Mruga{\l}a}\ and\ \citenamefont
  {Kraemer}(2005)}]{Mrugala_2005}%
  \BibitemOpen
  \bibfield  {author} {\bibinfo {author} {\bibfnamefont {F.}~\bibnamefont
  {Mruga{\l}a}}\ and\ \bibinfo {author} {\bibfnamefont {W.~P.}\ \bibnamefont
  {Kraemer}},\ }\href@noop {} {\bibfield  {journal} {\bibinfo  {journal} {J.\
  Chem.\ Phys.}\ }\textbf {\bibinfo {volume} {122}},\ \bibinfo {pages} {224321}
  (\bibinfo {year} {2005})}\BibitemShut {NoStop}%
\bibitem [{\citenamefont {Mruga{\l}a}\ and\ \citenamefont
  {Kraemer}(2013)}]{Mrugala_2015}%
  \BibitemOpen
  \bibfield  {author} {\bibinfo {author} {\bibfnamefont {F.}~\bibnamefont
  {Mruga{\l}a}}\ and\ \bibinfo {author} {\bibfnamefont {W.~P.}\ \bibnamefont
  {Kraemer}},\ }\href@noop {} {\bibfield  {journal} {\bibinfo  {journal} {J.\
  Chem.\ Phys.}\ }\textbf {\bibinfo {volume} {138}},\ \bibinfo {pages} {104315}
  (\bibinfo {year} {2013})}\BibitemShut {NoStop}%
\bibitem [{\citenamefont {Stoecklin}, \citenamefont {Lique},\ and\
  \citenamefont {Hochlaf}(2013)}]{Stoecklin}%
  \BibitemOpen
  \bibfield  {author} {\bibinfo {author} {\bibfnamefont {T.}~\bibnamefont
  {Stoecklin}}, \bibinfo {author} {\bibfnamefont {F.}~\bibnamefont {Lique}}, \
  and\ \bibinfo {author} {\bibfnamefont {M.}~\bibnamefont {Hochlaf}},\
  }\href@noop {} {\bibfield  {journal} {\bibinfo  {journal} {Pys.\ Chem.\
  Chem.\ Phys.}\ }\textbf {\bibinfo {volume} {15}},\ \bibinfo {pages} {13818}
  (\bibinfo {year} {2013})}\BibitemShut {NoStop}%
\bibitem [{\citenamefont {Ayouz}\ \emph {et~al.}(2011)\citenamefont {Ayouz},
  \citenamefont {Lopes}, \citenamefont {Raoult}, \citenamefont {Dulieu},\ and\
  \citenamefont {Kokoouline}}]{Ayouz}%
  \BibitemOpen
  \bibfield  {author} {\bibinfo {author} {\bibfnamefont {M.}~\bibnamefont
  {Ayouz}}, \bibinfo {author} {\bibfnamefont {R.}~\bibnamefont {Lopes}},
  \bibinfo {author} {\bibfnamefont {M.}~\bibnamefont {Raoult}}, \bibinfo
  {author} {\bibfnamefont {O.}~\bibnamefont {Dulieu}}, \ and\ \bibinfo {author}
  {\bibfnamefont {V.}~\bibnamefont {Kokoouline}},\ }\href@noop {} {\bibfield
  {journal} {\bibinfo  {journal} {Pys.\ Rev.\ A.}\ }\textbf {\bibinfo {volume}
  {83}},\ \bibinfo {pages} {052712} (\bibinfo {year} {2011})}\BibitemShut
  {NoStop}%
\bibitem [{\citenamefont {Talbi}\ and\ \citenamefont
  {Bacchus-Montabonel}(2010)}]{Talbi}%
  \BibitemOpen
  \bibfield  {author} {\bibinfo {author} {\bibfnamefont {D.}~\bibnamefont
  {Talbi}}\ and\ \bibinfo {author} {\bibfnamefont {M.~C.}\ \bibnamefont
  {Bacchus-Montabonel}},\ }\href@noop {} {\bibfield  {journal} {\bibinfo
  {journal} {Chem.\ Phys.\ Lett.}\ }\textbf {\bibinfo {volume} {485}},\
  \bibinfo {pages} {56} (\bibinfo {year} {2010})}\BibitemShut {NoStop}%
\bibitem [{\citenamefont {Gustafsson}(2013)}]{Classical}%
  \BibitemOpen
  \bibfield  {author} {\bibinfo {author} {\bibfnamefont {M.}~\bibnamefont
  {Gustafsson}},\ }\href@noop {} {\bibfield  {journal} {\bibinfo  {journal}
  {J.\ Chem.\ Phys.}\ }\textbf {\bibinfo {volume} {138}},\ \bibinfo {pages}
  {074308} (\bibinfo {year} {2013})}\BibitemShut {NoStop}%
\bibitem [{\citenamefont {Szab\'o}\ and\ \citenamefont
  {Lendvay}(2015{\natexlab{a}})}]{Szabo_1}%
  \BibitemOpen
  \bibfield  {author} {\bibinfo {author} {\bibfnamefont {P.}~\bibnamefont
  {Szab\'o}}\ and\ \bibinfo {author} {\bibfnamefont {G.}~\bibnamefont
  {Lendvay}},\ }\href@noop {} {\bibfield  {journal} {\bibinfo  {journal} {J.\
  Phys.\ Chem.\ A.}\ }\textbf {\bibinfo {volume} {119}},\ \bibinfo {pages}
  {7180} (\bibinfo {year} {2015}{\natexlab{a}})}\BibitemShut {NoStop}%
\bibitem [{\citenamefont {Szab\'o}\ and\ \citenamefont
  {Lendvay}(2015{\natexlab{b}})}]{Szabo_2}%
  \BibitemOpen
  \bibfield  {author} {\bibinfo {author} {\bibfnamefont {P.}~\bibnamefont
  {Szab\'o}}\ and\ \bibinfo {author} {\bibfnamefont {G.}~\bibnamefont
  {Lendvay}},\ }\href@noop {} {\bibfield  {journal} {\bibinfo  {journal} {J.\
  Phys.\ Chem.\ A.}\ }\textbf {\bibinfo {volume} {119}},\ \bibinfo {pages}
  {12485} (\bibinfo {year} {2015}{\natexlab{b}})}\BibitemShut {NoStop}%
\bibitem [{\citenamefont {de~Oliveira-Filho}, \citenamefont {Ornellas},\ and\
  \citenamefont {Bowman}(2014)}]{HBrOH}%
  \BibitemOpen
  \bibfield  {author} {\bibinfo {author} {\bibfnamefont {A.~G.~S.}\
  \bibnamefont {de~Oliveira-Filho}}, \bibinfo {author} {\bibfnamefont {F.~N.}\
  \bibnamefont {Ornellas}}, \ and\ \bibinfo {author} {\bibfnamefont {J.~M.}\
  \bibnamefont {Bowman}},\ }\href@noop {} {\bibfield  {journal} {\bibinfo
  {journal} {J.\ Phys.\ Chem.\ A.}\ }\textbf {\bibinfo {volume} {118}},\
  \bibinfo {pages} {12080} (\bibinfo {year} {2014})}\BibitemShut {NoStop}%
\bibitem [{\citenamefont {Herbst}\ and\ \citenamefont
  {Dunbar}(1991)}]{Herbst_1}%
  \BibitemOpen
  \bibfield  {author} {\bibinfo {author} {\bibfnamefont {E.}~\bibnamefont
  {Herbst}}\ and\ \bibinfo {author} {\bibfnamefont {R.~C.}\ \bibnamefont
  {Dunbar}},\ }\href@noop {} {\bibfield  {journal} {\bibinfo  {journal} {Mon.\
  Not.\ R.\ Astron.\ Soc.}\ }\textbf {\bibinfo {volume} {253}},\ \bibinfo
  {pages} {341} (\bibinfo {year} {1991})}\BibitemShut {NoStop}%
\bibitem [{\citenamefont {Herbst}(1985{\natexlab{a}})}]{Herbst_2}%
  \BibitemOpen
  \bibfield  {author} {\bibinfo {author} {\bibfnamefont {E.}~\bibnamefont
  {Herbst}},\ }\href@noop {} {\bibfield  {journal} {\bibinfo  {journal}
  {Astron.\ Atrophys.}\ }\textbf {\bibinfo {volume} {153}},\ \bibinfo {pages}
  {151} (\bibinfo {year} {1985}{\natexlab{a}})}\BibitemShut {NoStop}%
\bibitem [{\citenamefont {Herbst}(1985{\natexlab{b}})}]{Herbst_3}%
  \BibitemOpen
  \bibfield  {author} {\bibinfo {author} {\bibfnamefont {E.}~\bibnamefont
  {Herbst}},\ }\href@noop {} {\bibfield  {journal} {\bibinfo  {journal}
  {Astrphys.\ Jour.}\ }\textbf {\bibinfo {volume} {292}},\ \bibinfo {pages}
  {484} (\bibinfo {year} {1985}{\natexlab{b}})}\BibitemShut {NoStop}%
\bibitem [{\citenamefont {Cui}\ and\ \citenamefont {Thiel}(2014)}]{Thiel}%
  \BibitemOpen
  \bibfield  {author} {\bibinfo {author} {\bibfnamefont {G.}~\bibnamefont
  {Cui}}\ and\ \bibinfo {author} {\bibfnamefont {W.}~\bibnamefont {Thiel}},\
  }\href@noop {} {\bibfield  {journal} {\bibinfo  {journal} {J.\ Chem.\ Phys.}\
  }\textbf {\bibinfo {volume} {141}},\ \bibinfo {pages} {124101} (\bibinfo
  {year} {2014})}\BibitemShut {NoStop}%
\bibitem [{\citenamefont {Barbatti}(2011)}]{Barbatti}%
  \BibitemOpen
  \bibfield  {author} {\bibinfo {author} {\bibfnamefont {M.}~\bibnamefont
  {Barbatti}},\ }\href@noop {} {\bibfield  {journal} {\bibinfo  {journal}
  {Adv.\ Rev.}\ }\textbf {\bibinfo {volume} {1}},\ \bibinfo {pages} {620}
  (\bibinfo {year} {2011})}\BibitemShut {NoStop}%
\bibitem [{\citenamefont {Richter}\ \emph {et~al.}(2017)\citenamefont
  {Richter}, \citenamefont {Marquetand}, \citenamefont {Gonz\'alez-Vazquez},
  \citenamefont {Sola},\ and\ \citenamefont {Gonz\'alez}}]{SHARC}%
  \BibitemOpen
  \bibfield  {author} {\bibinfo {author} {\bibfnamefont {M.}~\bibnamefont
  {Richter}}, \bibinfo {author} {\bibfnamefont {P.}~\bibnamefont {Marquetand}},
  \bibinfo {author} {\bibfnamefont {J.}~\bibnamefont {Gonz\'alez-Vazquez}},
  \bibinfo {author} {\bibfnamefont {I.}~\bibnamefont {Sola}}, \ and\ \bibinfo
  {author} {\bibfnamefont {L.}~\bibnamefont {Gonz\'alez}},\ }\href@noop {}
  {\bibfield  {journal} {\bibinfo  {journal} {Chem.\ Phys.}\ }\textbf {\bibinfo
  {volume} {482}},\ \bibinfo {pages} {9} (\bibinfo {year} {2017})}\BibitemShut
  {NoStop}%
\bibitem [{\citenamefont {Du}\ and\ \citenamefont {Lan}(2014)}]{JADE}%
  \BibitemOpen
  \bibfield  {author} {\bibinfo {author} {\bibfnamefont {L.}~\bibnamefont
  {Du}}\ and\ \bibinfo {author} {\bibfnamefont {Z.}~\bibnamefont {Lan}},\
  }\href@noop {} {\bibfield  {journal} {\bibinfo  {journal} {J.\ Chem.\ Theo.\
  Comp.}\ }\textbf {\bibinfo {volume} {11}},\ \bibinfo {pages} {1360} (\bibinfo
  {year} {2014})}\BibitemShut {NoStop}%
\bibitem [{\citenamefont {Subotnik}\ \emph {et~al.}(2016)\citenamefont
  {Subotnik}, \citenamefont {Jain}, \citenamefont {Landry}, \citenamefont
  {Petit}, \citenamefont {Ouyang},\ and\ \citenamefont
  {Bellonzi}}]{Subotnik_rev}%
  \BibitemOpen
  \bibfield  {author} {\bibinfo {author} {\bibfnamefont {J.~E.}\ \bibnamefont
  {Subotnik}}, \bibinfo {author} {\bibfnamefont {E.}~\bibnamefont {Jain}},
  \bibinfo {author} {\bibfnamefont {B.}~\bibnamefont {Landry}}, \bibinfo
  {author} {\bibfnamefont {A.}~\bibnamefont {Petit}}, \bibinfo {author}
  {\bibfnamefont {W.}~\bibnamefont {Ouyang}}, \ and\ \bibinfo {author}
  {\bibfnamefont {N.}~\bibnamefont {Bellonzi}},\ }\href@noop {} {\bibfield
  {journal} {\bibinfo  {journal} {Ann.\ Rev.\ Phys.\ Chem.}\ }\textbf {\bibinfo
  {volume} {67}},\ \bibinfo {pages} {387} (\bibinfo {year} {2016})}\BibitemShut
  {NoStop}%
\bibitem [{\citenamefont {Petit}\ and\ \citenamefont
  {Subotnik}(2014)}]{Subotnik_1}%
  \BibitemOpen
  \bibfield  {author} {\bibinfo {author} {\bibfnamefont {A.~S.}\ \bibnamefont
  {Petit}}\ and\ \bibinfo {author} {\bibfnamefont {J.~E.}\ \bibnamefont
  {Subotnik}},\ }\href@noop {} {\bibfield  {journal} {\bibinfo  {journal} {J.\
  Chem.\ Phys.}\ }\textbf {\bibinfo {volume} {141}},\ \bibinfo {pages} {154108}
  (\bibinfo {year} {2014})}\BibitemShut {NoStop}%
\bibitem [{\citenamefont {Petit}\ and\ \citenamefont
  {Subotnik}(2015)}]{Subotnik_2}%
  \BibitemOpen
  \bibfield  {author} {\bibinfo {author} {\bibfnamefont {A.~S.}\ \bibnamefont
  {Petit}}\ and\ \bibinfo {author} {\bibfnamefont {J.~E.}\ \bibnamefont
  {Subotnik}},\ }\href@noop {} {\bibfield  {journal} {\bibinfo  {journal} {J.\
  Chem.\ Theo.\ Comp.}\ }\textbf {\bibinfo {volume} {11}},\ \bibinfo {pages}
  {4328} (\bibinfo {year} {2015})}\BibitemShut {NoStop}%
\bibitem [{\citenamefont {Mitric}, \citenamefont {Petersen},\ and\
  \citenamefont {Bonacic-Koutecky}(2009)}]{FISH}%
  \BibitemOpen
  \bibfield  {author} {\bibinfo {author} {\bibfnamefont {R.}~\bibnamefont
  {Mitric}}, \bibinfo {author} {\bibfnamefont {J.}~\bibnamefont {Petersen}}, \
  and\ \bibinfo {author} {\bibfnamefont {V.}~\bibnamefont {Bonacic-Koutecky}},\
  }\href@noop {} {\bibfield  {journal} {\bibinfo  {journal} {Phys.\ Rev.\ A.}\
  }\textbf {\bibinfo {volume} {79}},\ \bibinfo {pages} {053416} (\bibinfo
  {year} {2009})}\BibitemShut {NoStop}%
\bibitem [{\citenamefont {Mitric}\ \emph
  {et~al.}(2011{\natexlab{a}})\citenamefont {Mitric}, \citenamefont {Petersen},
  \citenamefont {Wohlgemuth}, \citenamefont {Werner},\ and\ \citenamefont
  {Bonacic-Koutecky}}]{Mitric_1}%
  \BibitemOpen
  \bibfield  {author} {\bibinfo {author} {\bibfnamefont {R.}~\bibnamefont
  {Mitric}}, \bibinfo {author} {\bibfnamefont {J.}~\bibnamefont {Petersen}},
  \bibinfo {author} {\bibfnamefont {M.}~\bibnamefont {Wohlgemuth}}, \bibinfo
  {author} {\bibfnamefont {U.}~\bibnamefont {Werner}}, \ and\ \bibinfo {author}
  {\bibfnamefont {V.}~\bibnamefont {Bonacic-Koutecky}},\ }\href@noop {}
  {\bibfield  {journal} {\bibinfo  {journal} {Phys.\ Chem.\ Chem.\ Phys.}\
  }\textbf {\bibinfo {volume} {13}},\ \bibinfo {pages} {8690} (\bibinfo {year}
  {2011}{\natexlab{a}})}\BibitemShut {NoStop}%
\bibitem [{\citenamefont {Lisinetskaya}\ and\ \citenamefont
  {Mitric}(2011)}]{Mitric_2}%
  \BibitemOpen
  \bibfield  {author} {\bibinfo {author} {\bibfnamefont {P.~G.}\ \bibnamefont
  {Lisinetskaya}}\ and\ \bibinfo {author} {\bibfnamefont {R.}~\bibnamefont
  {Mitric}},\ }\href@noop {} {\bibfield  {journal} {\bibinfo  {journal} {Phys.\
  Rev.\ A.}\ }\textbf {\bibinfo {volume} {88}},\ \bibinfo {pages} {033408}
  (\bibinfo {year} {2011})}\BibitemShut {NoStop}%
\bibitem [{\citenamefont {Mitric}\ \emph
  {et~al.}(2011{\natexlab{b}})\citenamefont {Mitric}, \citenamefont {Petersen},
  \citenamefont {Wohlgemuth}, \citenamefont {Werner}, \citenamefont
  {Bonacic-Koutecky}, \citenamefont {Woste},\ and\ \citenamefont
  {Jortner}}]{Mitric_3}%
  \BibitemOpen
  \bibfield  {author} {\bibinfo {author} {\bibfnamefont {R.}~\bibnamefont
  {Mitric}}, \bibinfo {author} {\bibfnamefont {J.}~\bibnamefont {Petersen}},
  \bibinfo {author} {\bibfnamefont {M.}~\bibnamefont {Wohlgemuth}}, \bibinfo
  {author} {\bibfnamefont {U.}~\bibnamefont {Werner}}, \bibinfo {author}
  {\bibfnamefont {V.}~\bibnamefont {Bonacic-Koutecky}}, \bibinfo {author}
  {\bibfnamefont {L.}~\bibnamefont {Woste}}, \ and\ \bibinfo {author}
  {\bibfnamefont {J.}~\bibnamefont {Jortner}},\ }\href@noop {} {\bibfield
  {journal} {\bibinfo  {journal} {J.\ Phys.\ Chem.\ A.}\ }\textbf {\bibinfo
  {volume} {115}},\ \bibinfo {pages} {3755} (\bibinfo {year}
  {2011}{\natexlab{b}})}\BibitemShut {NoStop}%
\bibitem [{\citenamefont {Humeniuk}\ and\ \citenamefont
  {Mitric}(2016)}]{Mitric_4}%
  \BibitemOpen
  \bibfield  {author} {\bibinfo {author} {\bibfnamefont {A.}~\bibnamefont
  {Humeniuk}}\ and\ \bibinfo {author} {\bibfnamefont {R.}~\bibnamefont
  {Mitric}},\ }\href@noop {} {\bibfield  {journal} {\bibinfo  {journal} {J.\
  Chem.\ Phys.}\ }\textbf {\bibinfo {volume} {144}},\ \bibinfo {pages} {234108}
  (\bibinfo {year} {2016})}\BibitemShut {NoStop}%
\bibitem [{\citenamefont {Mai}\ \emph {et~al.}(2011)\citenamefont {Mai},
  \citenamefont {Richter}, \citenamefont {Marquetand},\ and\ \citenamefont
  {Gonz\'alez}}]{Gonzalez_1}%
  \BibitemOpen
  \bibfield  {author} {\bibinfo {author} {\bibfnamefont {S.}~\bibnamefont
  {Mai}}, \bibinfo {author} {\bibfnamefont {M.}~\bibnamefont {Richter}},
  \bibinfo {author} {\bibfnamefont {P.}~\bibnamefont {Marquetand}}, \ and\
  \bibinfo {author} {\bibfnamefont {L.}~\bibnamefont {Gonz\'alez}},\
  }\href@noop {} {\bibfield  {journal} {\bibinfo  {journal} {J.\ Chem.\ Theo.\
  Comp.}\ }\textbf {\bibinfo {volume} {7}},\ \bibinfo {pages} {1253} (\bibinfo
  {year} {2011})}\BibitemShut {NoStop}%
\bibitem [{\citenamefont {Borin}\ \emph {et~al.}(2017)\citenamefont {Borin},
  \citenamefont {Mai}, \citenamefont {Marquetand},\ and\ \citenamefont
  {Gonz\'alez}}]{Gonzalez_2}%
  \BibitemOpen
  \bibfield  {author} {\bibinfo {author} {\bibfnamefont {A.}~\bibnamefont
  {Borin}}, \bibinfo {author} {\bibfnamefont {S.}~\bibnamefont {Mai}}, \bibinfo
  {author} {\bibfnamefont {P.}~\bibnamefont {Marquetand}}, \ and\ \bibinfo
  {author} {\bibfnamefont {L.}~\bibnamefont {Gonz\'alez}},\ }\href@noop {}
  {\bibfield  {journal} {\bibinfo  {journal} {Phys.\ Chem.\ Chem.\ Phys.}\
  }\textbf {\bibinfo {volume} {19}},\ \bibinfo {pages} {5888} (\bibinfo {year}
  {2017})}\BibitemShut {NoStop}%
\bibitem [{\citenamefont {Wang}\ \emph {et~al.}(2016)\citenamefont {Wang},
  \citenamefont {Liu}, \citenamefont {Cui}, \citenamefont {Fang},\ and\
  \citenamefont {Thiel}}]{Thiel_2}%
  \BibitemOpen
  \bibfield  {author} {\bibinfo {author} {\bibfnamefont {Y.~T.}\ \bibnamefont
  {Wang}}, \bibinfo {author} {\bibfnamefont {X.~Y.}\ \bibnamefont {Liu}},
  \bibinfo {author} {\bibfnamefont {G.}~\bibnamefont {Cui}}, \bibinfo {author}
  {\bibfnamefont {W.~H.}\ \bibnamefont {Fang}}, \ and\ \bibinfo {author}
  {\bibfnamefont {W.}~\bibnamefont {Thiel}},\ }\href@noop {} {\bibfield
  {journal} {\bibinfo  {journal} {Ang.\ Chem.\ Int.\ Ed.}\ }\textbf {\bibinfo
  {volume} {55}},\ \bibinfo {pages} {14009} (\bibinfo {year}
  {2016})}\BibitemShut {NoStop}%
\bibitem [{\citenamefont {Spörkel}\ and\ \citenamefont
  {Thiel}(2016)}]{Thiel_3}%
  \BibitemOpen
  \bibfield  {author} {\bibinfo {author} {\bibfnamefont {L.}~\bibnamefont
  {Spörkel}}\ and\ \bibinfo {author} {\bibfnamefont {W.}~\bibnamefont
  {Thiel}},\ }\href@noop {} {\bibfield  {journal} {\bibinfo  {journal} {J.\
  Chem.\ Phys.}\ }\textbf {\bibinfo {volume} {144}},\ \bibinfo {pages} {194108}
  (\bibinfo {year} {2016})}\BibitemShut {NoStop}%
\bibitem [{\citenamefont {Fazzi}, \citenamefont {Barbatti},\ and\ \citenamefont
  {Thiel}(2016)}]{Barbatti_2}%
  \BibitemOpen
  \bibfield  {author} {\bibinfo {author} {\bibfnamefont {D.}~\bibnamefont
  {Fazzi}}, \bibinfo {author} {\bibfnamefont {M.}~\bibnamefont {Barbatti}}, \
  and\ \bibinfo {author} {\bibfnamefont {W.}~\bibnamefont {Thiel}},\
  }\href@noop {} {\bibfield  {journal} {\bibinfo  {journal} {J.\ Am.\ Chem.\
  Soc.}\ }\textbf {\bibinfo {volume} {138}},\ \bibinfo {pages} {4502} (\bibinfo
  {year} {2016})}\BibitemShut {NoStop}%
\bibitem [{\citenamefont {Bjerre}\ and\ \citenamefont
  {Nikitin}(1967)}]{Surfhopp_original_1}%
  \BibitemOpen
  \bibfield  {author} {\bibinfo {author} {\bibfnamefont {A.}~\bibnamefont
  {Bjerre}}\ and\ \bibinfo {author} {\bibfnamefont {E.~E.}\ \bibnamefont
  {Nikitin}},\ }\href@noop {} {\bibfield  {journal} {\bibinfo  {journal}
  {Chem.\ Phys.\ Lett.}\ }\textbf {\bibinfo {volume} {1}},\ \bibinfo {pages}
  {179} (\bibinfo {year} {1967})}\BibitemShut {NoStop}%
\bibitem [{\citenamefont {Herman}(1984{\natexlab{a}})}]{Surfhopp_original_2}%
  \BibitemOpen
  \bibfield  {author} {\bibinfo {author} {\bibfnamefont {M.~F.}\ \bibnamefont
  {Herman}},\ }\href@noop {} {\bibfield  {journal} {\bibinfo  {journal} {J.\
  Chem.\ Phys.}\ }\textbf {\bibinfo {volume} {81}},\ \bibinfo {pages} {754}
  (\bibinfo {year} {1984}{\natexlab{a}})}\BibitemShut {NoStop}%
\bibitem [{\citenamefont {Herman}(1984{\natexlab{b}})}]{Surfhopp_original_3}%
  \BibitemOpen
  \bibfield  {author} {\bibinfo {author} {\bibfnamefont {M.~F.}\ \bibnamefont
  {Herman}},\ }\href@noop {} {\bibfield  {journal} {\bibinfo  {journal} {J.\
  Chem.\ Phys.}\ }\textbf {\bibinfo {volume} {81}},\ \bibinfo {pages} {764}
  (\bibinfo {year} {1984}{\natexlab{b}})}\BibitemShut {NoStop}%
\bibitem [{\citenamefont {Tully}(1990)}]{Surfhopp_original_4}%
  \BibitemOpen
  \bibfield  {author} {\bibinfo {author} {\bibfnamefont {J.~C.}\ \bibnamefont
  {Tully}},\ }\href@noop {} {\bibfield  {journal} {\bibinfo  {journal} {J.
  Chem.\ Phys.}\ }\textbf {\bibinfo {volume} {93}},\ \bibinfo {pages} {1061}
  (\bibinfo {year} {1990})}\BibitemShut {NoStop}%
\bibitem [{\citenamefont {Hammes-Schiffer}\ and\ \citenamefont
  {Tully}(1994)}]{Surfhopp_original_5}%
  \BibitemOpen
  \bibfield  {author} {\bibinfo {author} {\bibfnamefont {S.}~\bibnamefont
  {Hammes-Schiffer}}\ and\ \bibinfo {author} {\bibfnamefont {J.~C.}\
  \bibnamefont {Tully}},\ }\href@noop {} {\bibfield  {journal} {\bibinfo
  {journal} {J.\ Chem.\ Phys.}\ }\textbf {\bibinfo {volume} {101}},\ \bibinfo
  {pages} {4657} (\bibinfo {year} {1994})}\BibitemShut {NoStop}%
\bibitem [{\citenamefont {Coker}\ and\ \citenamefont
  {Xiao}(1994)}]{Surfhopp_original_6}%
  \BibitemOpen
  \bibfield  {author} {\bibinfo {author} {\bibfnamefont {D.~F.}\ \bibnamefont
  {Coker}}\ and\ \bibinfo {author} {\bibfnamefont {L.}~\bibnamefont {Xiao}},\
  }\href@noop {} {\bibfield  {journal} {\bibinfo  {journal} {J.\ Chem.\ Phys.}\
  }\textbf {\bibinfo {volume} {102}},\ \bibinfo {pages} {496} (\bibinfo {year}
  {1994})}\BibitemShut {NoStop}%
\bibitem [{\citenamefont {Ruckenbauer}\ \emph {et~al.}(2016)\citenamefont
  {Ruckenbauer}, \citenamefont {Mai}, \citenamefont {Marquetand},\ and\
  \citenamefont {Gonz\'alez}}]{Gonzalez_3}%
  \BibitemOpen
  \bibfield  {author} {\bibinfo {author} {\bibfnamefont {M.}~\bibnamefont
  {Ruckenbauer}}, \bibinfo {author} {\bibfnamefont {S.}~\bibnamefont {Mai}},
  \bibinfo {author} {\bibfnamefont {P.}~\bibnamefont {Marquetand}}, \ and\
  \bibinfo {author} {\bibfnamefont {L.}~\bibnamefont {Gonz\'alez}},\
  }\href@noop {} {\bibfield  {journal} {\bibinfo  {journal} {J.\ Chem.\ Phys.}\
  }\textbf {\bibinfo {volume} {144}},\ \bibinfo {pages} {074303} (\bibinfo
  {year} {2016})}\BibitemShut {NoStop}%
\bibitem [{\citenamefont {Davydov}(1976)}]{Davydov}%
  \BibitemOpen
  \bibfield  {author} {\bibinfo {author} {\bibfnamefont {A.~S.}\ \bibnamefont
  {Davydov}},\ }\href@noop {} {\emph {\bibinfo {title} {Quantum Mechanics}}}\
  (\bibinfo  {publisher} {Pergamon Press, Oxford},\ \bibinfo {year}
  {1976})\BibitemShut {NoStop}%
\bibitem [{\citenamefont {Schatz}\ and\ \citenamefont
  {Ratner}(2002)}]{SchatzRatner}%
  \BibitemOpen
  \bibfield  {author} {\bibinfo {author} {\bibfnamefont {G.~C.}\ \bibnamefont
  {Schatz}}\ and\ \bibinfo {author} {\bibfnamefont {M.~A.}\ \bibnamefont
  {Ratner}},\ }\href@noop {} {\emph {\bibinfo {title} {Quantum Mechanics in
  Chemitry}}}\ (\bibinfo  {publisher} {Dover Publications New York},\ \bibinfo
  {year} {2002})\BibitemShut {NoStop}%
\bibitem [{\citenamefont {Noda}\ and\ \citenamefont
  {Zare}(1982)}]{Franck-Condon}%
  \BibitemOpen
  \bibfield  {author} {\bibinfo {author} {\bibfnamefont {C.}~\bibnamefont
  {Noda}}\ and\ \bibinfo {author} {\bibfnamefont {R.~N.}\ \bibnamefont
  {Zare}},\ }\href@noop {} {\bibfield  {journal} {\bibinfo  {journal} {J.\
  Mol.\ Spectrosc.}\ }\textbf {\bibinfo {volume} {95}},\ \bibinfo {pages} {254}
  (\bibinfo {year} {1982})}\BibitemShut {NoStop}%
\bibitem [{\citenamefont {Babb}\ and\ \citenamefont
  {Dalgarno}(1995)}]{Dalgarno}%
  \BibitemOpen
  \bibfield  {author} {\bibinfo {author} {\bibfnamefont {J.~F.}\ \bibnamefont
  {Babb}}\ and\ \bibinfo {author} {\bibfnamefont {A.}~\bibnamefont
  {Dalgarno}},\ }\href@noop {} {\bibfield  {journal} {\bibinfo  {journal}
  {Phys.\ Rev.\ A.}\ }\textbf {\bibinfo {volume} {51}},\ \bibinfo {pages}
  {3021} (\bibinfo {year} {1995})}\BibitemShut {NoStop}%
\bibitem [{\citenamefont {Kramers}\ and\ \citenamefont {ter
  Haar}(1946)}]{Kramers}%
  \BibitemOpen
  \bibfield  {author} {\bibinfo {author} {\bibfnamefont {H.~A.}\ \bibnamefont
  {Kramers}}\ and\ \bibinfo {author} {\bibfnamefont {D.}~\bibnamefont {ter
  Haar}},\ }\href@noop {} {\bibfield  {journal} {\bibinfo  {journal} {Bull.\
  Astronom.\ Inst.\ Netherl.}\ }\textbf {\bibinfo {volume} {10}},\ \bibinfo
  {pages} {137} (\bibinfo {year} {1946})}\BibitemShut {NoStop}%
\bibitem [{\citenamefont {Brown}\ and\ \citenamefont
  {Carrington}(2003)}]{Semiquant}%
  \BibitemOpen
  \bibfield  {author} {\bibinfo {author} {\bibfnamefont {J.~M.}\ \bibnamefont
  {Brown}}\ and\ \bibinfo {author} {\bibfnamefont {A.}~\bibnamefont
  {Carrington}},\ }\href@noop {} {\emph {\bibinfo {title} {Rotational
  Spectroscopy of Diatomic Molecules}}}\ (\bibinfo  {publisher} {Cambridge
  University Press},\ \bibinfo {year} {2003})\BibitemShut {NoStop}%
\bibitem [{\citenamefont {Qin}, \citenamefont {Wang},\ and\ \citenamefont
  {Zhang}(1992)}]{Sympfour}%
  \BibitemOpen
  \bibfield  {author} {\bibinfo {author} {\bibfnamefont {M.}~\bibnamefont
  {Qin}}, \bibinfo {author} {\bibfnamefont {D.}~\bibnamefont {Wang}}, \ and\
  \bibinfo {author} {\bibfnamefont {M.}~\bibnamefont {Zhang}},\ }\href@noop {}
  {\bibfield  {journal} {\bibinfo  {journal} {J.\ Comp.\ Math.}\ }\textbf
  {\bibinfo {volume} {9}},\ \bibinfo {pages} {211} (\bibinfo {year}
  {1992})}\BibitemShut {NoStop}%
\bibitem [{\citenamefont {Kramida}\ \emph {et~al.}(2015)\citenamefont
  {Kramida}, \citenamefont {{Yu.~Ralchenko}}, \citenamefont {Reader},\ and\
  \citenamefont {{NIST ASD Team}}}]{NIST_ASD}%
  \BibitemOpen
  \bibfield  {author} {\bibinfo {author} {\bibfnamefont {A.}~\bibnamefont
  {Kramida}}, \bibinfo {author} {\bibnamefont {{Yu.~Ralchenko}}}, \bibinfo
  {author} {\bibfnamefont {J.}~\bibnamefont {Reader}}, \ and\ \bibinfo {author}
  {\bibnamefont {{NIST ASD Team}}},\ }\href@noop {} {}\bibinfo {howpublished}
  {{NIST Atomic Spectra Database (ver. 5.3), [Online]. Available:
  {\tt{http://physics.nist.gov/asd}} [2017, April 25]. National Institute of
  Standards and Technology, Gaithersburg, MD.}} (\bibinfo {year}
  {2015})\BibitemShut {NoStop}%
\end{thebibliography}%

\end{document}